\documentclass[a4paper,fleqn]{cas-sc}

\usepackage[numbers]{natbib}
\usepackage{subfig}
\usepackage{amsmath}
\usepackage{graphicx}
\usepackage{amssymb,amsfonts,bm}
\usepackage{algorithmic}
\usepackage{textcomp}
\usepackage{xcolor}
\usepackage{tabularx}
\usepackage[switch]{lineno}
\def\tsc#1{\csdef{#1}{\textsc{\lowercase{#1}}\xspace}}
\tsc{WGM}
\tsc{QE}
\begin{document}
\let\WriteBookmarks\relax
\def\floatpagepagefraction{1}
\def\textpagefraction{.001}
%\let\printorcid\relax % 可去掉页面下方的ORCID(s)

% \shorttitle{<short title of the paper for running head>}    
\shorttitle{Future Full-Ocean Deep SSPs Prediction based on H-LSTM Neural Networks}   

% \shortauthors{<short author list for running head>} 
\shortauthors{Jiajun Lu et al.}

\title[mode = title]{Future Full-Ocean Deep SSPs Prediction based on Hierarchical Long Short-Term Memory Neural Networks}  

% \tnotemark[<tnote number>] 
\tnotemark[1]

\tnotetext[1]{This document is the results of the research project funded by Natural Science Foundation of Shandong Province (ZR2023QF128), China Postdoctoral Science Foundation (2022M722990), Qingdao Postdoctoral Science Foundation (QDBSH20220202061), National Natural Science Foundation of China (NSFC:62271459), National Defense Science and Technology Innovation Special Zone Project: Marine Science and Technology Collaborative Innovation Center (22-05-CXZX-04-01-02), and the Fundamental Research Funds for the Central Universities, Ocean University of China (202313036).}

\author[1]{Jiajun Lu}
\ead{lujiajun@stu.ouc.edu.cn} 
\credit{Methodology, Software, Writing - Original draft preparation}

\author[1]{Hao Zhang}
\ead{zhanghao@ouc.edu.cn}
\credit{Conceptualization, Writing - editing and review}

\author[1]{Pengfei Wu}
\ead{wupengfei@stu.ouc.edu.cn} 
\credit{Software, Writing - editing and review}

\author[1]{Sijia Li}
\ead{lisijia9430@stu.ouc.edu.cn} 
\credit{Software, Writing - editing and review}

\author[1]{Wei Huang}[orcid=0000-0002-8284-2310]
\ead{hw@ouc.edu.cn} 
\credit{Conceptualization of this study, Methodology, Software, Writing - editing and review}
\cormark[1]

\address[1]{Ocean University of China, Qingdao 266100, China}
\cortext[1]{Corresponding author} 


%\linenumbers
\begin{abstract}
The spatial-temporal distribution of underwater sound velocity affects the propagation mode of underwater acoustic signals. Therefore, rapid estimation and prediction of underwater sound velocity distribution is crucial for providing underwater positioning, navigation and timing (PNT) services. Currently, sound speed profile (SSP) inversion methods have a faster time response rate compared to direct measurement methods, however, most SSP inversion methods focus on constructing spatial dimensional sound velocity fields and are highly dependent on sonar observation data, thus high requirements have been placed on observation data sources. To explore the distribution pattern of sound velocity in the time dimension and achieve future SSP prediction without sonar observation data, we propose a hierarchical long short-term memory (H-LSTM) neural network for SSP prediction. By our SSP prediction method, the sound speed distribution could be estimated without any on-site data measurement process, so that the time efficiency could be greatly improved. Through comparing with other state-of-the-art methods, H-LSTM has better accuracy performance on prediction of monthly average sound velocity distribution, which is less than 1 m/s in different depth layers.
\end{abstract}


\begin{highlights}
\item Predicting sound speed by hierarchical long short-term memory neural networks.
\item No on-site data measurement is required.
\item A rolling window helps fully use of training data to reduce over-fitting.
\item Capturing the hierarchical periodic features of sound speed profiles.
\item Root mean square error of sound speed predicting is less than 1 m/s.

\end{highlights}

% Keywords

\maketitle

% Main text
\section{Introduction}
Underwater sound speed distribution is one of the most important parameters for underwater positioning, navigation and timing (PNT) system, because it affects the propagation mode of underwater acoustic signals according to \cite{Wang2013Marine}. With the increasing demand for precision performance in PNT, it is necessary to quickly and accurately obtain the regional sound speed distribution, and even predict the future sound speed distribution.

\indent The acquisition of ocean sound speed profile (SSP) mainly includes SSP measurement method and SSP inversion method. The SSP can be measured directly by using a sound velocity profiler (SVP) \cite{Xu2012Research}, or can be measured indirectly by using a conductivity, temperature and depth profiler (CTD) \cite{Yuan2009Technical,Wang2014SBE,SeaSun2023} or an expendable CTD profiler (XCTD) \cite{XCTD2023} combined with empirical sound speed formulas. However, the measurement of SSP by SVP, CTD or XCTD usually takes a long time. For example, it takes at least 80 minutes to measure an SSP with the depth of 2000 meters by SVP or CTD according to \cite{Huang2023Fast}. Although the SSP measurement time can be significantly shortened by using the XCTD, which is about 20 minutes for the same 2000 meters depth range, it is still not real-time and the depth range is limited by the sensor’s pressure resistance constraints.

\indent In recent years, ocean SSPs inversion methods have been widely studied to fast obtain sound speed distribution. The traditional methods for SSP inversion are mainly divided into three categories: matched field processing, compressed sensing, and deep learning methods. Most of the SSP inversion methods focus on spatial SSP construction, and most of them rely on real-time sonar observation data and related data such as temperature and salinity. There are relatively few studies on SSP prediction.

\indent In 1979, Munk and Wunsch first proposed the concept of SSP in \cite{Munk1979Ocean,Munk1983Ocean}, and put forward the idea of inverting the SSP, using the sound speed propagation time as the basis information for inverting the SSP. In 1995, Tolstoy proposed a matching field processing framework for SSP inversion in \cite{tolstoy1991acoustic}, where an effective solution is provided that avoid the difficult work of establishing a mapping relationship from sound field distribution to SSPs. In 1997, Taroudakis combined the matching field processing with the modal phase inversion in \cite{Taroudakis1997On}. In 2000, Shen et al. demonstrated the feasibility of inversion of SSPs by using empirical orthogonal functions (EOFs) in shallow sea area \cite{Shen1999Feasiblity}. In 2002, Zhang et al. proposed an SSP inversion method based on matched beam processing in \cite{zhang2002beam}, which can reconstruct shallow water SSPs. However, these methods did not gain good SSP inversion accuracy. 

\indent To improve the accuracy performance, Han et al. proposed an improved SSPs estimation method based on the traditional EOF in \cite{Han2009Marine}. Zhang proposed an inversion method based on 3D spatial characteristic sound ray searching and sound propagation time calculation model in \cite{zhang2012Inversion,Zhang2012InversionO}. Zhang et al. proposed a method for constructing a time-varying model of regional small time-scale SSP based on layered EOFs in \cite{Zhang2022Modeling}. We previously proposed a deep-learning model for SSP inversion in \cite{Huang2019Collaborating}. The above methods have effectively improved the inversion accuracy of SSPs, but they rely on real-time sonar observation data during the inversion process, so that reduces the time efficiency. Recently, Li et al. proposed a self-organizing map (SOM) neural network that combines surface sound speed for SSP estimation in \cite{Li2022Inversion}, which achieves the construction of SSP without sound field observation data, but there is still a lot of room for improvement in accuracy. Moreover, it is unable to predict future sound speed distribution.

\indent The above various inversion methods can achieve relatively accurate spatial dimension SSPs inversion, but they lack the ability to capture the changing law of sound velocity distribution with time. To tackle the problem of SSP prediction in the time dimension, we propose a future sound speed prediction method based on hierarchical long short-term memory (H-LSTM) neural networks. The historical SSP data in the spatial area where the prediction tasks are located are adopted as references, among which historical SSP distribution information sorted in chronological order is used to accurately predict future sound speed distribution in the target area.

\indent The contribution of this paper is that we propose an H-LSTM method for the rapid prediction of the future sound speed distribution. This method realizes the accurate prediction of the sound speed distribution of the full-ocean deep in the future referring to the historical SSP data. By our SSP prediction method, the sound speed distribution could be estimated without any on-site data measurement process, so that the time efficiency could be greatly improved. We first process the historical sound speed distribution data in layers and set up different H-LSTM neural network models for different depth layers, then train and predict the sound speed value in different depth layers, finally combine the prediction results of each H-LSTM model's layer to form full-ocean deep SSP.

\indent The rest of this paper is organized as follows: Section 2 proposes the H-LSTM prediction method, and then introduces the specific implementation details of the H-LSTM prediction experiment. Section 3 gives the experimental results and discussions, fully verifies the H-LSTM model, and compares the performance with other state-of-the-art methods. Finally, conclusions are given in Section 4.

\section{Methodology}
Considering the seasonal variation of temperature that affects sound velocity and the inconsistent variation patterns at different depths, we propose a H-LSTM model for predicting sound speed based on depth stratification. In this section, we will first introduce the classic LSTM model in the following text, and then introduce our proposed H-LSTM model.
\subsection{Structure of LSTM}
\subsubsection{LSTM}
Long short-term memory (LSTM) neural network is a special type of recurrent neural network that can solve the problems of gradient vanishing and explosion in recurrent neural networks. LSTM performs well in time series prediction problems \cite{Hochreiter1997LSTM}.

\begin{figure}[h]
	\includegraphics[width=\linewidth]{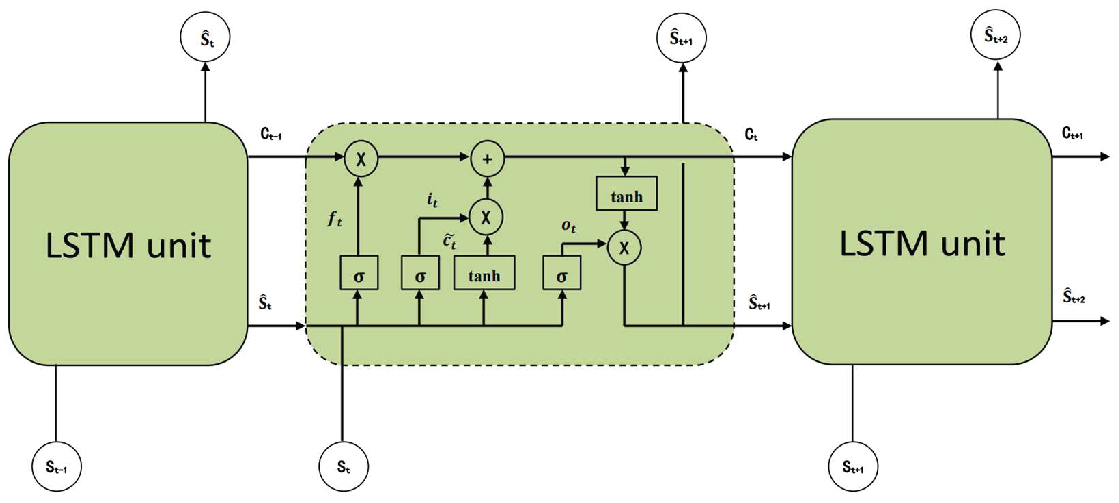}
	\caption{Unit structure of LSTM.\label{fig01}}
\end{figure}

\indent LSTM controls the transmission of information by introducing a gating mechanism, which includes input gates, forgetting gates, and output gates. There is an internal state called "cell state" in LSTM that can read, write, and clear information at different time steps of the sequence. The forgetting gate determines how much information in the previous cell state $C_{t-1}$ needs to be discarded, and how much can be retained in the current cell state $C_{t}$. The input gate determines how much information of currently input is saved in the current cell state $C_{t}$. The output gate determines how much information in the cell state $C_{t}$ can be output to the estimated value $h_t$ for the next moment. The unit structure of LSTM is shown in figure \ref{fig01}.

\begin{figure}[h]
	\centering
	\includegraphics[width=0.6\linewidth]{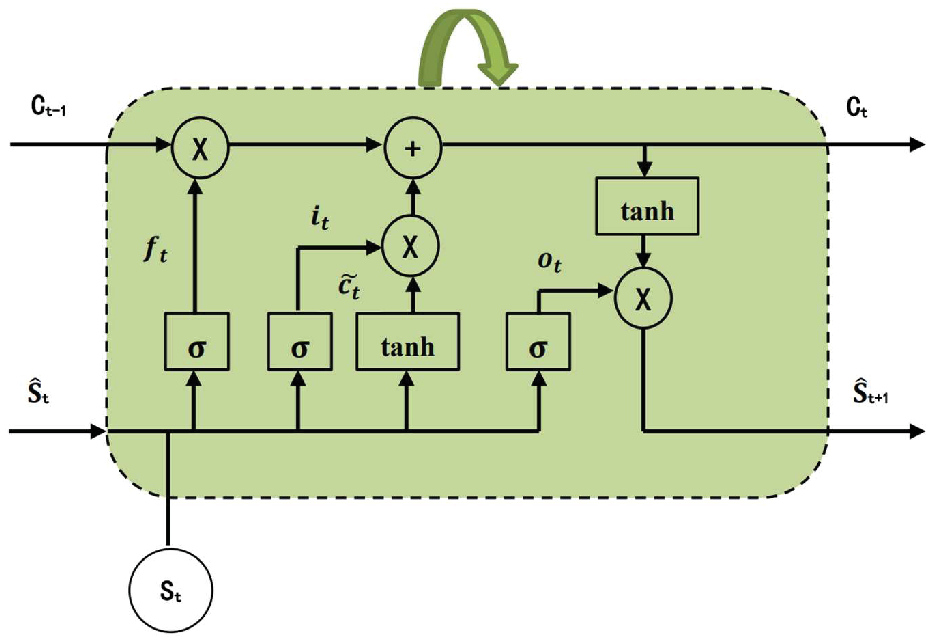}
	\caption{Folding-formed unit structure of LSTM.\label{fig02}}
\end{figure}

For simplicity, we will fold multiple LSTM units for representation, as shown in figure \ref{fig02}. The LSTM neural network achieves long-term data information storage by updating the state of internal gates. In the figure, ``X'' and ``+'' represent multiplication and addition operations, respectively, while LSTM unit represents the same internal structure as the intermediate unit. $\sigma$ and $tanh$ denote the sigmoid activation function and Tanh activation function, respectively. The t-1 moment represents the previous moment, the t moment represents the current moment, and the t+1 moment represents the next moment. Thus, $S_{t-1},\quad S_{t},\quad S_{t+1}$ represent the SSP input values from time moment t-1 to t+1 respectively. $C_{t-1},\quad C_{t},\quad C_{t+1}$ represent the memory units at time moment t-1, t, and t+1, respectively. $\hat{S}_{t},\quad \hat{S}_{t+1},\quad \hat{S}_{t+2}$ represent the estimated SSP value of the hidden layer output from time moment t-1 to t+1, respectively. W and b are weight matrices and biases corresponding to the three types of gates.

\indent The calculation process of LSTM neural network mainly consists of the following four steps.
\vspace{-2mm}
\paragraph{1) Calculating output $f_t$ of forgetting gate:}
\vspace{-2mm}
\begin{equation}
	f_t = \sigma(W_f \cdot [\hat{S}_t,S_t] + b_f),
\end{equation}
where $W_f$ represents the weight matrix of the forgetting gate, $b_f$ represents the bias of the forgetting gate.
\vspace{-2mm}
\paragraph{2) Calculating output $i_t$ of input gate and candidate cell state $\tilde{C}_t$:}
\vspace{-2mm}
\begin{equation}
	i_t = \sigma(W_i \cdot [\hat{S}_t,S_t] + b_i),
\end{equation}
\begin{equation}
	\tilde{C}_t = tanh(W_c \cdot [\hat{S}_t,S_t] + b_c),
\end{equation}
where $W_i$ and $W_c$ represent the weight matrices of the input gate and candidate cell state inputs, respectively. $b_i$ and $b_c$ represent the biases of the input gate and candidate cell state inputs, respectively.
\vspace{-2mm}
\paragraph{3) Updating cell state $C_t$:}
\vspace{-2mm}
\begin{equation}
	\tilde{C}_t = f_t \cdot C_{t-1} + i_t \cdot \tilde{C}_t.
\end{equation}
\vspace{-2mm}
\paragraph{4) Predicting $o_t$ of output gate and output $\hat{S}_{t+1}$ of hidden layer:}
\vspace{-2mm}
\begin{equation}
	o_t = \sigma(W_{o}\cdot [\hat{S}_t,S_t] + b_o),
\end{equation}
\begin{equation}
	\hat{S}_{t+1} = o_t \cdot tanh(C_t),
\end{equation}
where $W_o$ represents the weight matrix of the output gate, $b_o$ represents the bias of the output gate.

\subsubsection{H-LSTM}
Although LSTM performs well in time series prediction models, it has a fatal drawback when processing spatio-temporal data. It must expand the input features into one-dimensional time series before processing, thus losing some corresponding feature information during the processing. 

Considering the different variations in sound velocity at different depths, we propose a hierarchical long short-term memory (H-LSTM) neural network model. The core idea of our method is to treat sound velocity distributions at different depths as multiple time varying sequences. Then, corresponding H-LSTM models are defined for each depth layer, and these models are trained separately to predict sound speed values at various depths at a certain moment. Final SSP is obtained through the combination of predict sound speed value at different depth layers.

\begin{figure}[h]
	\includegraphics[width=\linewidth]{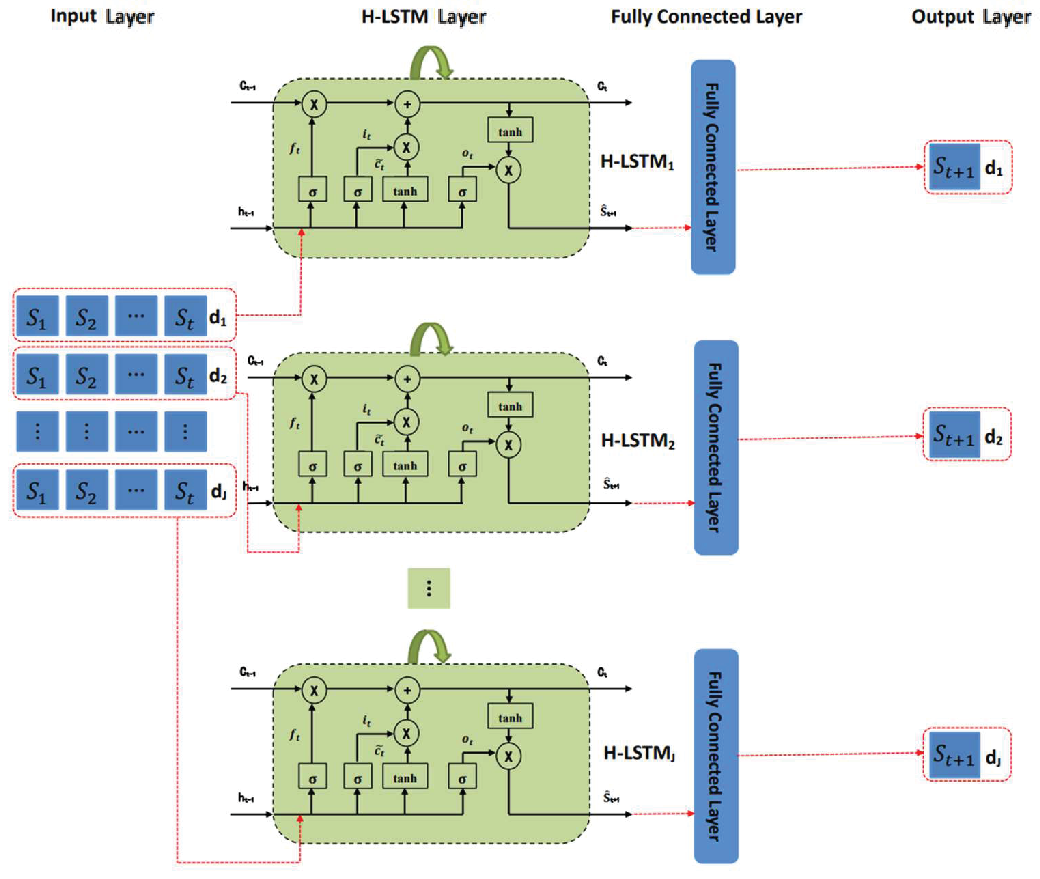}
	\caption{The structure of H-LSTM.\label{fig03}}
\end{figure}

In this paper, the H-LSTM model is mainly composed of an input layer, an H-LSTM layer, a fully connected layer, and an output layer. The model structure of H-LSTM is shown in Figure \ref{fig03}, where $d_J$ represents the $J$th depth layer, and $H-LSTM_J$ represents the H-LSTM model defined for the $J$th layer. Before inputting the data, the data is first layered according to different depth layers, and then normalized into hierarchical standardized SSP data. The input hierarchical standardized SSP data is processed in the H-LSTM layer, retaining important information and extracting the relationships between SSP data at different times. The purpose of adding a fully connected layer between the H-LSTM layer and the output layer is to improve the fitting performance of the model, and finally, the output layer will offer future SSP prediction results.
\begin{figure}[h]
	\centering
	\includegraphics[width=0.6\linewidth]{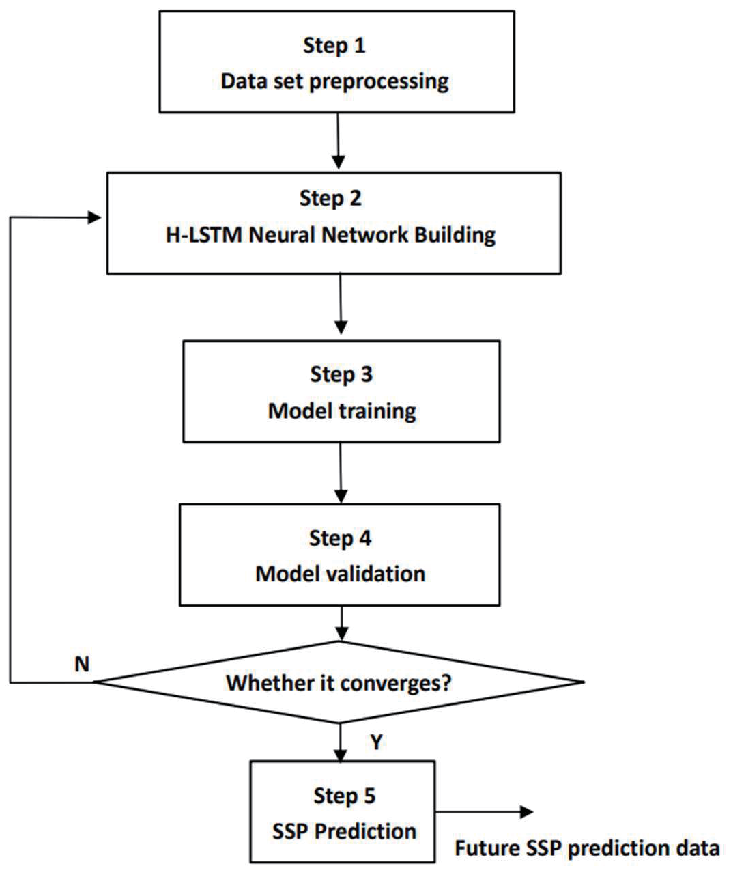}
	\caption{Flow chart of H-LSTM for SSP prediction.\label{fig04}}
\end{figure}
\subsection{Flowchart of H-LSTM for SSP Prediction}
The specific implementation steps of the sound speed distribution prediction method based on hierarchical long short-term memory (H-LSTM) neural network are shown in Figure \ref{fig04}, including data set preprocessing, H-LSTM neural network building, model training, model validation and SSP prediction.

\subsubsection{Dataset pre-processing}
\paragraph{1) Data layering by depth}
Due to different variations in sound velocity at different depths, data need to be layered first before feeding into H-LSTM. Firstly, the SSP data is subjected to a standardized linear layering process in depth direction, with a total of J layers divided. Then the data are sorted in the time dimension, with a monthly time resolution.

\indent The standardized dataset of historical hierarchical sound speed distribution is represented as:
\begin{equation}\label{eq07}
	\mathbf{S}={\mathbf{S}_1,\mathbf{S}_2,...,\mathbf{S}_i},i=1,2,...,I,
\end{equation}
where $\mathbf{S}_i$ represents the $i$th SSP. $\mathbf{S}_i$ can be expressed in detail as:
\begin{equation}\label{eq08}
	\mathbf{S}_i=\{s_{i,d_1},s_{i,d_2},...,s_{i,d_j}\}^T,j=0,1,...,J,
\end{equation}
where $s_{i,d_j}$ is the sound speed value of the $i$th SSP at the $j$th depth layer. The subscript $d_j$ represents the $j$th depth layer. The $\mathbf{S}$ dataset in equation (\ref{eq07}) is arranged in chronological order, with a monthly time resolution. In equation (\ref{eq08}), j is the depth layer index label, where the maximum common depth layer for the distribution of sound speed in the entire sea is $d_J$. Then, $\mathbf{S}$ can be arranged in chronological order with a monthly time resolution as:
\begin{equation}
	\mathbf{S}=\left[\begin{split}
		s_{1,d_1}\quad 	s_{2,d_1} \quad ... \quad s_{i,d_1}\\
		s_{1,d_2}\quad 	s_{2,d_2} \quad ... \quad s_{i,d_2}\\
		...\quad\quad 	... \quad\quad ... \quad\quad ...\\
		s_{1,d_j}\quad 	s_{2,d_j} \quad ... \quad s_{i,d_j}\\
	\end{split}\right].
\end{equation}

\paragraph{2) Data segmentation}
To evaluate and optimize the performance of the H-LSTM neural network model, we divided the historical hierarchical sound speed distribution time standardized dataset $\mathbf{S}$ into training sets and validation sets.

\indent The time period of data change is determined by the objective changing law of data over time, where the time period of data change is taken as C. The selection of the training dataset is related to the start time of the SSP to be predicted. If the SSP at time t+1 is to be predicted, the training dataset is the SSP data from time t+1-nC to time t of $\mathbf{S}$. That is, the data from the first n consecutive cycles of the prediction start time is taken as the training dataset, and the SSP data at time t+1 is used as the validation dataset. The specific representations of the hierarchical training set T and the hierarchical validation set V are shown in formulas (\ref{eq10}) and (\ref{eq11}), respectively. (t>nC)

\begin{equation}\label{eq10}
	\mathbf{T}=\left[\begin{split}
		s_{t-nC+1,d_1}\quad 	s_{t-nC+2,d_1} \quad ... \quad s_{t,d_1}\\
		s_{t-nC+1,d_2}\quad 	s_{t-nC+2,d_2} \quad ... \quad s_{t,d_2}\\
		...\quad\quad\quad\quad 	... \quad\quad\quad ... \quad\quad ...\\
		s_{t-nC+1,d_j}\quad 	s_{t-nC+2,d_j} \quad ... \quad s_{t,d_j}\\
	\end{split}\right],
\end{equation}
\begin{equation}\label{eq11}
	\mathbf{V}=\left[\begin{split}
		s_{t+1,d_1}\\
		s_{t+1,d_2}\\
		...\\
		s_{t+1,d_j}\\
	\end{split}\right].
\end{equation}

\paragraph{3) Further processing of hierarchical training sets}
In order to accelerate the convergence speed and improve generalization ability of the model, we normalize the hierarchical training sets and further divide it into a training input set and a training output set.

\indent The hierarchical training dataset $\mathbf{T}$ is normalized to $\mathbf{\tilde{T}}$ as:
\begin{equation}\label{eq12}
	\mathbf{\tilde{T}}=\left[\begin{split}
		\tilde{s}_{t-nC+1,d_1}\quad 	\tilde{s}_{t-nC+2,d_1} \quad ... \quad \tilde{s}_{t,d_1}\\
		\tilde{s}_{t-nC+1,d_2}\quad 	\tilde{s}_{t-nC+2,d_2} \quad ... \quad \tilde{s}_{t,d_2}\\
		...\quad\quad\quad\quad 	... \quad\quad\quad ... \quad\quad ...\\
		\tilde{s}_{t-nC+1,d_j}\quad 	\tilde{s}_{t-nC+2,d_j} \quad ... \quad \tilde{s}_{t,d_j}\\
	\end{split}\right].
\end{equation}
Then, the normalized $\mathbf{\tilde{T}}$ is divided into hierarchical training input set $\mathbf{\tilde{T}}^{in}$ and hierarchical training output set $\mathbf{\tilde{T}}^{out}$. A staggered one-step training method is adopted for the training set, that is, using the $j$th column of data to predict the $j+1$th column. Hierarchical training input set $\mathbf{\tilde{T}}^{in}$ and hierarchical training output set $\mathbf{\tilde{T}}^{out}$ are given as:
\begin{equation}
	\mathbf{\tilde{T}}^{in}=\left[\begin{split}
		\tilde{s}_{t-nC+1,d_1}\quad 	\tilde{s}_{t-nC+2,d_1} \quad ... \quad \tilde{s}_{t-1,d_1}\\
		\tilde{s}_{t-nC+1,d_2}\quad 	\tilde{s}_{t-nC+2,d_2} \quad ... \quad \tilde{s}_{t-1,d_2}\\
		...\quad\quad\quad\quad 	... \quad\quad\quad ... \quad\quad ...\quad\\
		\tilde{s}_{t-nC+1,d_j}\quad 	\tilde{s}_{t-nC+2,d_j} \quad ... \quad \tilde{s}_{t-1,d_j}\\
	\end{split}\right],
\end{equation}
\begin{equation}
	\mathbf{\tilde{T}}^{out}=\left[\begin{split}
		\tilde{s}_{t-nC+2,d_1}\quad 	\tilde{s}_{t-nC+3,d_1} \quad ... \quad \tilde{s}_{t,d_1}\\
		\tilde{s}_{t-nC+2,d_2}\quad 	\tilde{s}_{t-nC+3,d_2} \quad ... \quad \tilde{s}_{t,d_2}\\
		...\quad\quad\quad\quad 	... \quad\quad\quad ... \quad\quad ...\quad\\
		\tilde{s}_{t-nC+2,d_j}\quad 	\tilde{s}_{t-nC+3,d_j} \quad ... \quad \tilde{s}_{t,d_j}\\
	\end{split}\right].
\end{equation}

\subsubsection{H-LSTM neural network building}
SSP prediction is executed in layers. An H-LSTM network is constructed for each depth layer, with an input layer in each layer. One H-LSTM layer is selected as the hidden layer, and the number of neurons in the hidden layer is N. A fully connected layer with a linear activation function is added between the hidden layer and the output layer. The model structure diagram has been provided in the H-LSTM principle introduction section.

\subsubsection{Model training}
During model training, H-LSTM networks at each depth layer are trained separately. The learning rate of each depth layer is set to be $\alpha_{d_j}$ and iteration number is set to be $m_{d_j}$, where $d_j$ represents the corresponding depth layer.

\indent When training the H-LSTM$_j, j = 1,2,...,J$ network in the first depth layer, the input is the time series $\mathbf{\tilde{T}}^{in}_{d_j}=\left[\tilde{s}_{t+1-nC,d_j},\tilde{s}_{t+2-nC,d_j},...,\tilde{s}_{t-1,d_j}\right]$, and the corresponding output of the model is the time series $\mathbf{\tilde{T}}^{out}_{d_j}=\left[\tilde{s}_{t+2-nC,d_j},\tilde{s}_{t+3-nC,d_j},...,\tilde{s}_{t,d_j}\right]$.

\subsubsection{Model validation}
\paragraph{1) Model output}
After the model is trained, the last column of the hierarchical training output set $\mathbf{\tilde{T}}^{out}$ is used as the input of the model to predict future SSP data. When conducting layered verification, taking the first depth layer as an example, the input of the model is $\mathbf{\tilde{T}}^{out}_{d_1}(end)=\tilde{s}_{t,d_1}$. The predicted first layer SSP data at the next moment is $P_{d_1} = \hat{s}_{t+1,d_1}$, where $\hat{s}_{t+1,d_1}$ is the predicted sound speed value of the first depth layer.

\indent Combining the prediction results of each layer into a hierarchical column vector, the final predicted hierarchical SSP is $\hat{\mathbf{P}}$:
\begin{equation}
	\hat{\mathbf{P}}=\left[\begin{split}
		\hat{s}_{t+1,d_1}\\
		\hat{s}_{t+1,d_2}\\
		...\\
		\hat{s}_{t+1,d_j}\\
	\end{split}\right],j=1,2,...,J.
\end{equation}

\paragraph{2) Output data denormalization}
Due to the normalization process performed on the training dataset before model training, the predicted sound speed data should be denormalized and restored to the scale range of the original data, making it easier to be compared and analyzed. The predicted result $\check{P}$ after denormalization processing is:
\begin{equation}
	\mathbf{\check{P}}=\left[\begin{split}
		\check{s}_{t+1,d_1}\\
		\check{s}_{t+1,d_2}\\
		...\\
		\check{s}_{t+1,d_j}\\
	\end{split}\right],j=1,2,...,J,
\end{equation}
where $\check{s}_{t+1,d_j}$ is the result of reverse-normalizing the sound speed value predicted in the $j$th depth layer.

\paragraph{3) Cost function}
The root mean square error (RMSE) between prediction SSP data and label SSP data is adopted as cost function:
\begin{equation}
	RMSE = \sqrt{\frac{\sum(\mathbf{\check{P}}-\mathbf{V})^2}{J}}.
\end{equation}

If there is no significant change in the RMSE value compared to the previous one, the process will go to step 5 according to figure \ref{fig04}, otherwise the process will return to step 2.

\subsubsection{SSP Prediction}
According to the specified prediction time resolution and start time of the prediction task, the number of prediction iterations and the historical hierarchical sound speed reference dataset corresponding to start time of the prediction are determined. We use the historical hierarchical sound speed reference dataset as the training set of the model, repeat the model output step in step 4 to obtain the prediction result, and then repeat the output data denormalization step in step 4 to obtain the predicted hierarchical SSP data. Finally, full-ocean deep interpolation on the predicted hierarchical data is performed to obtain future full-ocean deep SSP data.

\section{Results and Discussions}
\subsection{Experiment settings}
\subsubsection{Data source}
\begin{figure}[h]
	\includegraphics[width=\linewidth]{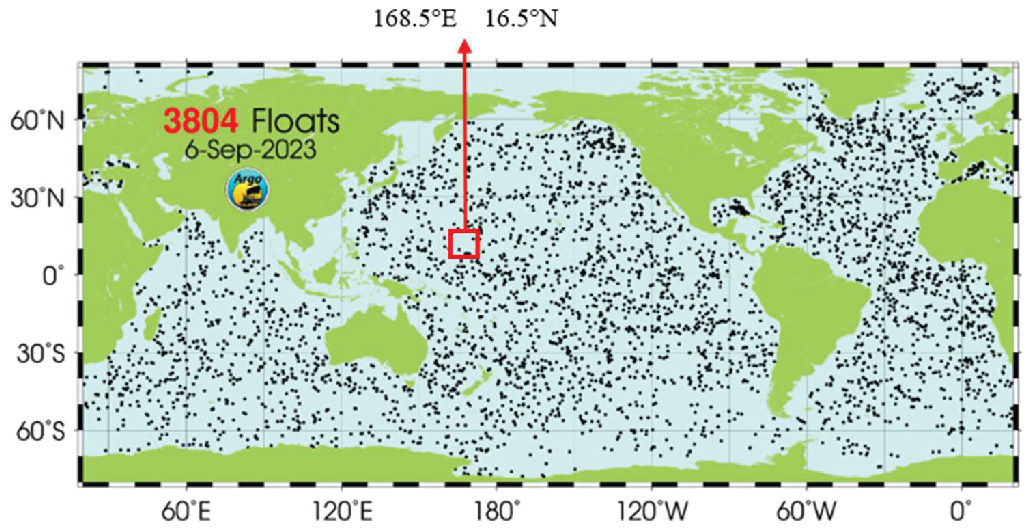}
	\caption{Spatial position of SSP samples.\label{fig05}}
\end{figure}

In order to verify the H-LSTM SSP prediction method proposed in this article, the region shown in figure \ref{fig05} was selected as the prediction location, specifically located in the central Pacific region at $168.5^\circ$ E and $16.5^\circ$ N.

\begin{table}[h] 
	\caption{Data information.\label{tab1}}
	\begin{tabularx}{\textwidth}{CCCCC}
		\toprule
		\textbf{Study Area}	& \textbf{Input} & \textbf{Time Dimension} & \textbf{Temporal Resolution} & \textbf{Spacial Resolution}\\
		\midrule
		$168.5^\circ E,16.5^\circ N$ & SSP & 2017-2021 & Mean of Month & $1^\circ$\\
		\bottomrule
	\end{tabularx}
\end{table}

In order to verify the performance of the proposed H-LSTM SSP prediction model, 60 months of measured SSP data from the global Argo dataset at this prediction location from 2017 to 2021 were selected to test the H-LSTM prediction model. This dataset is GDCSM-Argo dataset provided by the Global Ocean Argo System (Hangzhou) Observation and Research Station of the Ministry of Natural Resources \cite{Zhang2021Argo}. The GDCSM-Argo dataset is a global grid dataset with a spatial resolution of 1° and a monthly temporal resolution. Table \ref{tab1} provides the data information used in this section.

\subsubsection{Label data generation}
Before training the H-LSTM prediction model, the data needs to be subjected to standardized sequences in depth direction. According to the partitioning method introduced in Section 2, it is divided into 58 layers. The standardized linear layering is described as follows: data are divided into layers every 5 meters from 0 to 10 meters; divided into layers every 10 meters from 10 to 180 meters; divided into layers every 20 meters from 180 to 460 meters; divided into layers every 50 meters from 500 to 1250 meters; divided into layers every 100 meters from 1300 to 1900 meters. At a depth of over 1900 meters, it is divided into a total of 58 layers. Then the data are sorted in a time dimension, as the selected data are from 2017 to 2021 with a total of 60 months.

\indent The historical hierarchical sound speed distribution time standardized dataset $\mathbf{S}$ after deep stratification and time sorting processing can be represented as a sound speed value data matrix with 58 rows and 60 columns:
\begin{equation}
	\mathbf{S}=\left[\begin{split}
		s_{1,d_1}\quad 	s_{2,d_1} \quad ... \quad s_{60,d_1}\\
		s_{1,d_2}\quad 	s_{2,d_2} \quad ... \quad s_{60,d_2}\\
		...\quad\quad 	... \quad\quad ... \quad\quad ...\\
		s_{1,d_{58}}\quad 	s_{2,d_{58}} \quad ... \quad s_{60,d_{58}}\\
	\end{split}\right].
\end{equation}

\subsubsection{Platform and H-LSTM parameters}
The data processing involved in this study were all implemented in MATLAB R2021a, and the main parameter settings of the H-LSTM model in the experiment are shown in Table \ref{tab2}.

\begin{table}[H] 
	\caption{Key parameters of the H-LSTM model.\label{tab2}}
	\begin{tabularx}{\textwidth}{CC}
		\toprule
		\textbf{Key Parameters}	& \textbf{Settings} \\
		\midrule
		Layers & 4\\
		Hidden size & 128\\
		Learning rate & 0.01\\
		Epoch & 300\\
		Loss function & MSE\\
		\bottomrule
	\end{tabularx}
\end{table}

\subsection{Baselines}
To evaluate the predictive performance of the H-LSTM network model more intuitively, we conducted several experiments in contrast with the mean value prediction method, polynomial fitting method \cite{Liu2019Sound}, and BP neural network prediction method \cite{Yu2020SSP}.

\subsubsection{Mean value prediction}
The future sound speed distribution can be roughly estimated using the average of historical sound speed distributions with in a small scale of spatial area. Therefore, we use the average SSP of historical periods to replace the future SSP as a baseline.

\subsubsection{Polynomial fitting}
Liu et al. proposed an SSP prediction method based on polynomial fitting in \cite{Liu2019Sound}. The polynomial was used to fit the mean of the measured shallow water SSP, and extended in the deep sea to form a full-ocean deep SSP. The fitted SSP can be applied to the prediction of future SSP. The core idea is to use an n-orders polynomial to fit the SSP S(z), which can be expressed as:
\begin{equation}
	S(z) = \sum_{i=0}^{N}\frac{S^{(n)}(0)}{n!}z^i = \sum_{i=0}^{N}a_iz^i, a_i = \frac{S^{(n)}(0)}{n!},
\end{equation}
where $S^{(n)}(0)$ is the $n$th derivative of S(z), S is the value of sound speed, and z is the depth value. The polynomial coefficient vectors $[a_0,a_1,...,a_n]$ can describe the main characteristics of underwater SSPs. The polynomial fitting method here is also used as a comparative experiment for our proposed H-LSTM prediction method.

\subsubsection{BP neural network}
Yu et al. proposed an SSP inversion and prediction method based on back propagation (BP) neural networks in \cite{Yu2020SSP}. This method is mainly based on the nonlinear function approximation ability of BP neural networks, combining historical SSP data to construct an SSP prediction model.

\subsection{Effect of Training Dataset on SSP Prediction Performance}
To verify that the proposed H-LSTM model prediction method can accurately predict future SSPs with at least a certain amount of historical SSP information learned. We selected 1-year, 2-year, 3-year, and 4-year data from the time standardized dataset $\mathbf{S}$ formed by 60 months of measured SSP data from the prediction location from 2017 to 2021 as the learning samples of the model. We tested the predictive performance of the H-LSTM model under different learning samples, and compared the root mean square error (RMSE) between the predicted SSP data and the real SSP data. The comparison results are shown in Table \ref{tab3}.

\begin{table}[H] 
	\caption{Effect of training dataset on SSP prediction performance.\label{tab3}}
	\begin{tabularx}{\textwidth}{CCCCC}
		\toprule
		\textbf{Data set}	& \textbf{1 year} & \textbf{2 years} & \textbf{3 years} & \textbf{4 years}\\
		\midrule
		\textbf{RMSE(m/s)} & 1.0159 & 0.5785 &\textbf{0.4953} & 0.4934\\
		\bottomrule
	\end{tabularx}
\end{table}

Table \ref{tab3} takes the prediction of January 2021 SSP as an example, using historical SSP data from 1-4 years before the start of the prediction as the training set of the H-LSTM model. Analysis of the corresponding RMSE data shows that as the training set data increases, the predictive performance of the model gradually improves. If the training dataset is too small, the model cannot capture the temporal variation characteristics of the data. Thus, at least three or more complete cycles of historical SSP data for model training is necessary before the prediction start time to ensure accurate prediction performance.

\subsection{Accuracy performance of H-LSTM}

\begin{figure}[!htbp]
	\includegraphics[width=\linewidth]{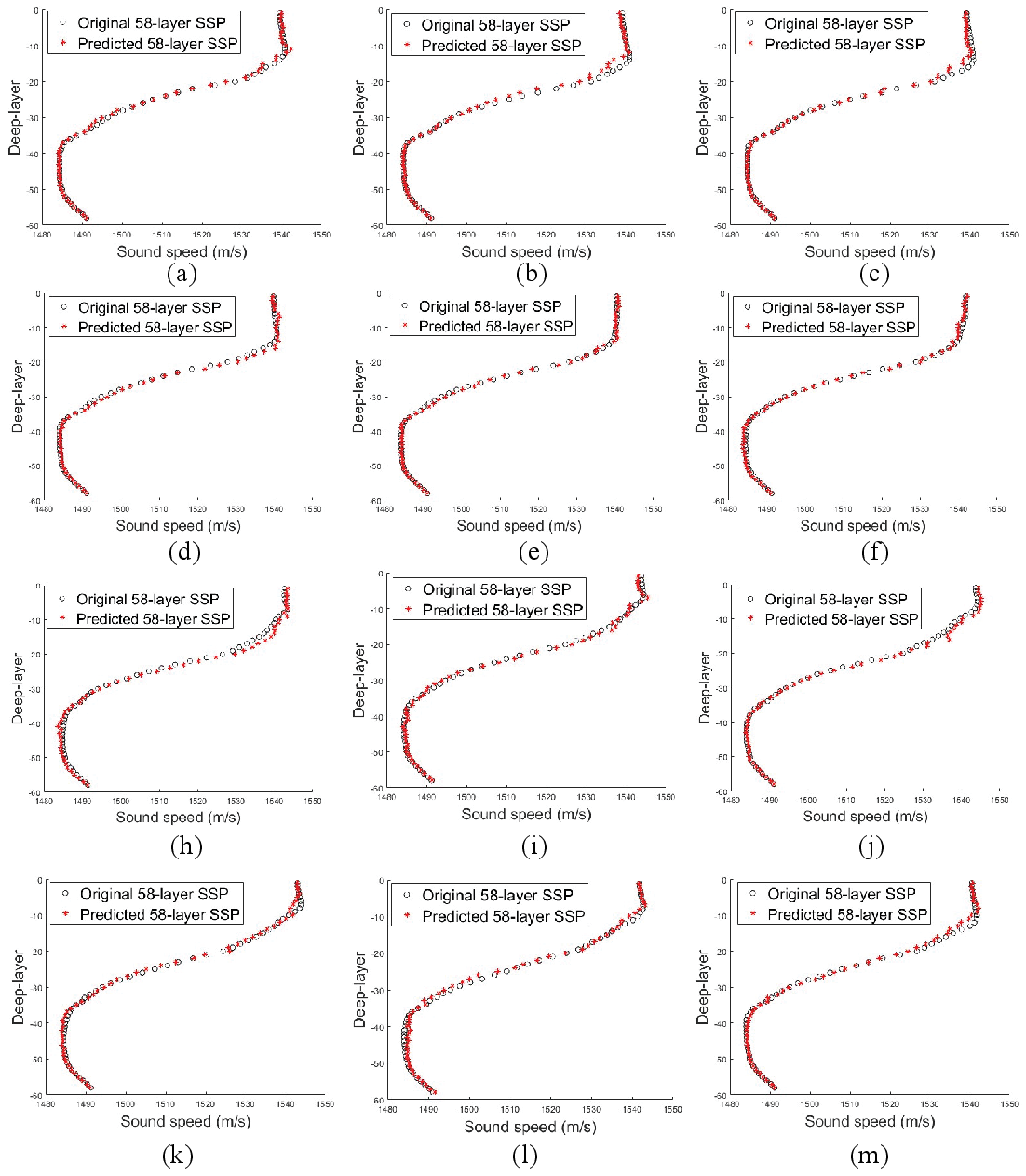}
	\caption{Comparison of actual 58-layer SSP and predicted 58-layer SSP for each month in 2021. (a)~(l) represent January to December, respectively.\label{fig06}}
\end{figure}

To verify the accuracy of the model's prediction of SSPs, sound speed distribution data from the first four complete cycles of the prediction start time were extracted from $\mathbf{S}$. Taking the prediction start time as October 2021 as an example, 48 historical hierarchical sound speed distribution data from October 2017 to September 2021 were used as learning samples for the model. The comparison between the original 58-layer sound speed data and the predicted 58-layer sound speed data for the 12 months of 2021 is shown in figure \ref{fig06}.

Through comparing the original 58-layer sound speed data with the predicted 58-layer sound speed data, it can be seen that the H-LSTM network model can make accurate predictions for future sound speed data, and the predicted sound speed values for each month in 2021 are similar to the actual sound speed values for each month in 2021.

\begin{figure}[!htbp]
	\includegraphics[width=\linewidth]{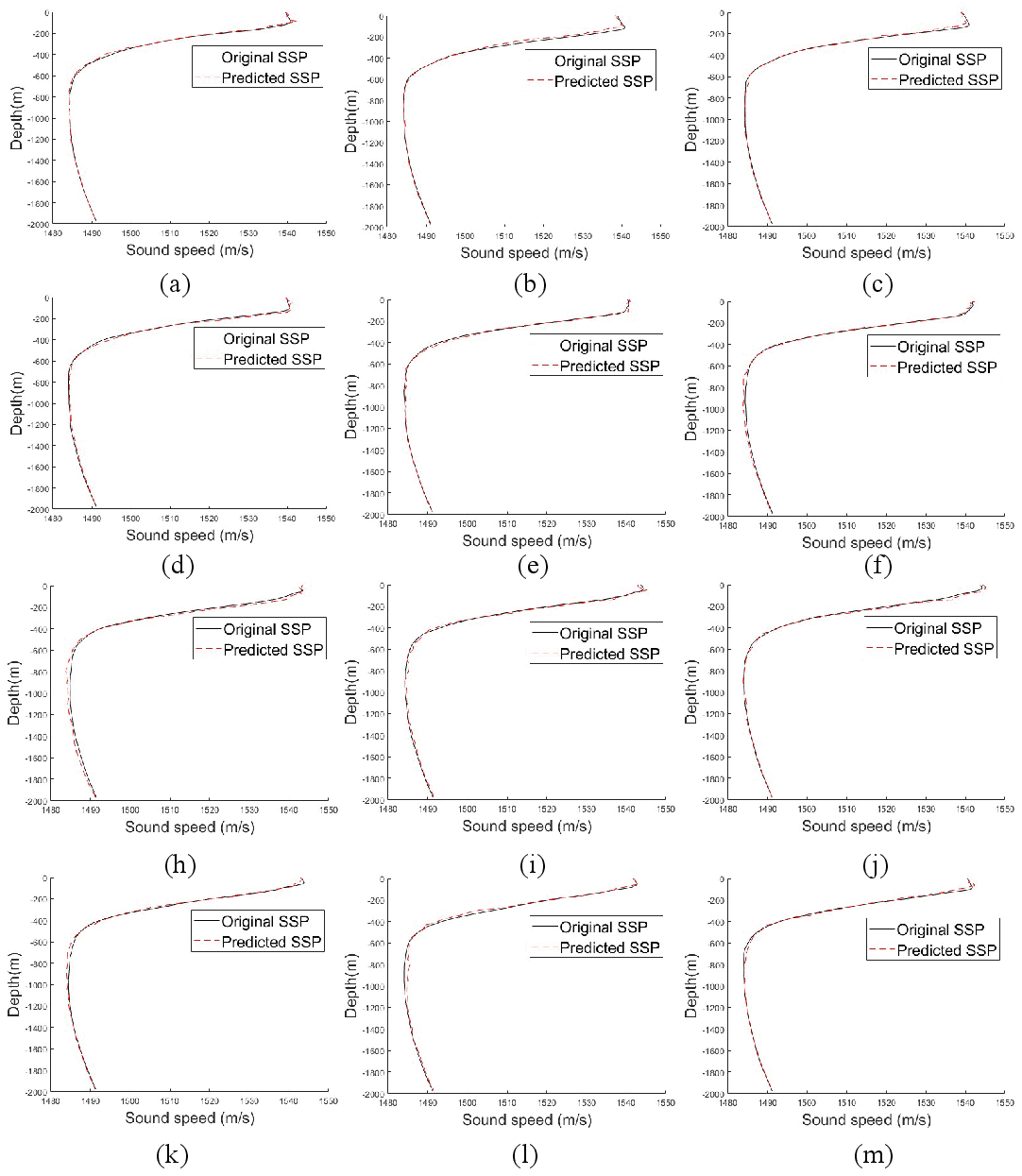}
	\caption{Comparison chart of predicted full-ocean deep SSP and original full-ocean deep SSP for each month in 2021. (a)~(l) represent January to December, respectively.\label{fig07}}
\end{figure}

Due to the fact that the time standardized dataset $\mathbf{S}$ is not full-ocean deep data, but data from 58 unequally spaced depth layers, the predicted results are also data from 58 depth layers. To test the predictive performance of the H-LSTM model for the full-ocean deep SSP, linear interpolation method was used to interpolate the predicted hierarchical SSP and the verified hierarchical SSP within the full ocean depth range. The full ocean depth here was taken as 0-1975 meters. The comparison between the predicted full ocean depth SSP and the actual full ocean depth SSP for each month in 2021 is given in Figure \ref{fig07}.

By comparing the predicted full ocean depth SSP for each month in 2021 with the real full ocean depth SSP, it can be seen that the H-LSTM network model can make accurate predictions of the future full ocean depth SSP, and the predicted future SSP can accurately represent the main characteristics of the actual SSP.

Table \ref{tab4} presents the prediction errors for each month in the next year, represented by the RMSE between the predicted full-ocean deep SSP and the real full-ocean deep SSP.

\begin{table}[!htbp] 
	\caption{Prediction errors for each month in the next year.\label{tab4}}
	\begin{tabularx}{\textwidth}{CC}
		\toprule
		\textbf{Predicted Area}	& \textbf{Pacific Ocean} \\
		\midrule
		\textbf{Depth}	& 0-1975 m \\
		\textbf{Year}	& 2021 \\
		\textbf{Month}	& RMSE (m/s) \\
		\textbf{1}	&  0.4934\\
		\textbf{2}	&  1.1310\\
		\textbf{3}	&  0.6630\\
		\textbf{4}	&  0.7163\\
		\textbf{5}	&  0.4586\\
		\textbf{6}	&  0.4618\\
		\textbf{7}	&  0.9990\\
		\textbf{8}	&  0.6212\\
		\textbf{9}	&  0.6955\\
		\textbf{10}	&  0.5565\\
		\textbf{11}	&  0.8606\\
		\textbf{12}	&  0.6443\\
		\textbf{Average Result}	& 0.6917 \\
		\bottomrule
	\end{tabularx}
\end{table}

\subsection{Performance comparison with other methods}
To fully validate the feasibility of our proposed H-LSTM neural network model for predicting future SSP, we conducted several experiments in contrast with the mean value prediction method, polynomial fitting method, and BP neural network prediction method.

In comparative experiments, taking the prediction of October 2021 as an example, the H-LSTM model used 48 layered SSPs data from October 2017 to September 2021 as a training set to predict future SSP. The mean value prediction method took the mean of four consecutive years of historical SSP data before October 2021 as the predicted future SSP data. The polynomial fitting method performed polynomial fitting on the historical SSP data of two consecutive years before October 2021 to obtain the predicted future SSP. There are two reasons for using the historical data of the two consecutive periods before the start time of the prediction as learning samples. Firstly, the polynomial fitting does not require too much data to achieve SSP prediction, and secondly, it is used to distinguish it from the mean value prediction method. The BP neural network used the same training dataset as H-LSTM as the model's training set to predict future SSP.

\begin{figure}[!htbp]
	\centering
	\includegraphics[width=0.6\linewidth]{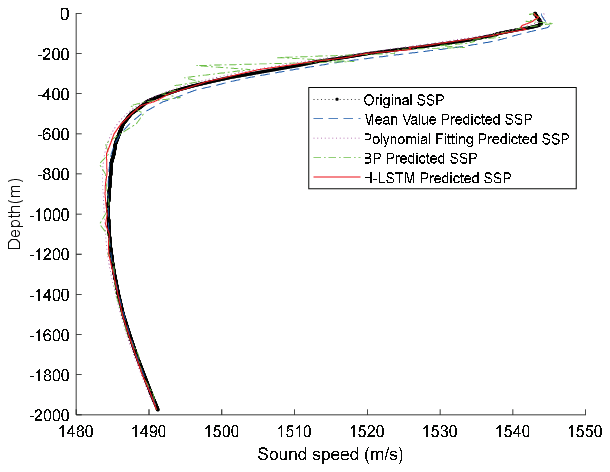}
	\caption{Comparison between predicted full-ocean deep SSP using four methods and original full-ocean deep SSP.\label{fig08}}
\end{figure}

To more intuitively compare the predicted full-ocean deep SSP at future times using several methods, we provided a comparison between the predicted full-ocean deep SSPs using these methods and the real full-ocean deep SSP. The comparison of the four experimental predicted SSPs is shown in figure \ref{fig08}. From figure \ref{fig08}, it can be seen that for the mean value prediction method, the SSP prediction effect is better in the deep-sea part (below 900 meters), but the prediction results are relatively poor in the part above 900 meters. For polynomial fitting, it can roughly fit the main features of future SSP smoothly, but there are some numerical deviations overall. For the BP neural network prediction method, its predicted future SSP is not smooth enough, fluctuating repeatedly around the real SSP, losing the main features of the actual SSP, and this result is not suitable as a prediction result for future SSP. For the H-LSTM neural network prediction method we proposed, the predicted full-ocean deep SSP does not lose the main features of the actual SSP compared with the real SSP, nor does it have an overall numerical deviation problem. In both shallow and deep-sea areas, the prediction results are relatively good, with only minor numerical deviations in some details.

\begin{figure}[!htbp]
	\includegraphics[width=\linewidth]{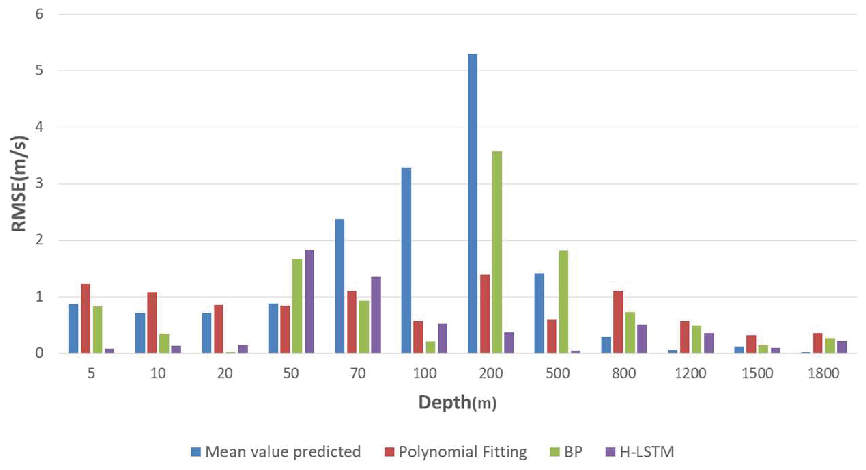}
	\caption{Comparison of four methods for predicting effects at different depths.\label{fig09}}
\end{figure}

In order to compare the performance of several methods in the future SSP prediction process at full-ocean deep more clearly, we selected three layers of data each from the surface layer, seasonal thermocline, main thermocline, and deep-sea isothermal layer of marine SSP to compare the errors between the predicted SSP and the actual SSP. The depths corresponding to the error data correspond to depths of 5 meters, 10 meters, 20 meters, 50 meters, 70 meters, 100 meters, 200 meters, 500 meters, 800 meters, 1200 meters, 1500 meters, and 1800 meters, respectively. They contain a typical four layers structure of SSP and roughly cover the full-ocean deep SSP data. The performance comparison of the four methods for predicting SSP is shown in figure \ref{fig09}.

\begin{table}[!htbp] 
	\caption{Comparison of RMSE of four prediction methods.\label{tab5}}
	\begin{tabularx}{\textwidth}{CCCCC}
		\toprule
		\textbf{Area}	& $168.5^\circ E,16.5^\circ N$ & $168.5^\circ E,16.5^\circ N$ & $168.5^\circ E,16.5^\circ N$ & $168.5^\circ E,16.5^\circ N$ \\
		\midrule
		\textbf{Predict Time} & 2021.10 & 2021.10 & 2021.10 & 2021.10 \\
		\textbf{Method} & Mean value & Polynomial fitting & BP Neural Network & \textbf{H-LSTM}\\
		\textbf{Dataset} & 2017-10-2021.09 & 2019-10-2021.09 & 2017-10-2021.09 & 2017-10-2021.09\\
		\textbf{RMSE (m/s)} & 1.5959 & 0.9548 & 1.7861 & \textbf{0.5565}\\
		\bottomrule
	\end{tabularx}
\end{table}

\begin{figure}[!htbp]
	\centering
	\includegraphics[width=0.6\linewidth]{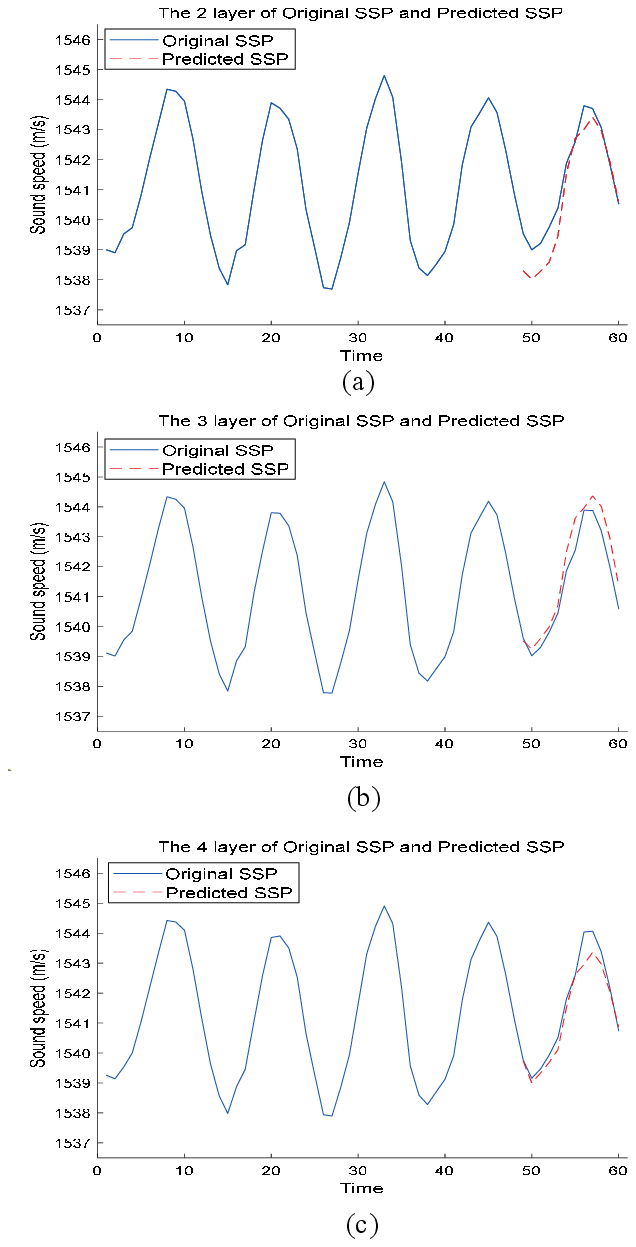}
	\caption{Comparison between the predicted trend and the original trend. (a) Ocean depth of 5 meters. (b) Ocean depth of 10 meters. (c) Ocean depth of 20 meters.\label{fig10}}
\end{figure}

From figure \ref{fig09}, it can be seen that for the mean value prediction method, its maximum prediction error occurs at a depth of 200 meters, exceeding 5 m/s. For polynomial fitting, the prediction error is relatively stable at each depth layer, with a maximum prediction error of no more than 1.5m/s. However, in most depth layers, the error exceeds 1m/s. For the BP neural network prediction method, its maximum prediction error also occurs at a depth of 200 meters, exceeding 3m/s. However, for the H-LSTM neural network prediction method we proposed, its prediction error is relatively small, with the maximum prediction error not exceeding 2m/s. Among the 12 selected layers, only two layers have a prediction error exceeding 1m/s, and the minimum prediction error is less than 0.5m/s. In summary, H-LSTM performs the best in the future SSP prediction process at full-ocean depth.

We provide a comparison of the overall RMSE results of four methods in predicting full-ocean deep SSP in Table \ref{tab5}. From the data in the table, it can be seen that H-LSTM predicts the most accurate full-ocean deep SSP, with a RMSE of 0.5565m/s.

\subsection{H-LSTM’s Performance on Predicting Cyclical Changes of SSPs}
To test the performance of the H-LSTM model in predicting periodic changes in SSP, we used historical SSPs data from 2017 to 2020 as learning samples and conducted a 12 steps prediction on future SSP data, namely predicting SSPs for all months of the next year. To ensure temporal rigor, as a single SSP data is consistent in time, the second layer data (corresponding to a depth of 5 meters), the third layer data (corresponding to a depth of 10 meters), and the fourth layer data (corresponding to a depth of 20 meters) are randomly selected between depth layers 1-58 to test whether this method can accurately capture the periodic changes in sound speed distribution. The schematic diagrams for comparing the periodic change trends of the predicted data in the second, third, and fourth depth layers with the original SSP data periodic change trends are shown in figures \ref{fig10}.

The 60 solid blue line data in the figure represent the real SSP data for the 60 months from 2017 to 2021 at the corresponding depth layer, with the first 48 being training data and the last 12 being validation data. The 12 red dashed line data represent the predicted SSP data for the next 12 months at the corresponding depth layer, which is compared with the validation data. It can be intuitively seen that our proposed H-LSTM method can accurately capture the periodic changes of SSP over time.

\section{Conclusions}

To satisfy the accuracy and time-efficient requirements of constructing underwater sound velocity field, we propose an H-LSTM method for future full-ocean deep SSP prediction. By our SSP prediction method, the sound speed distribution could be estimated without any on-site data measurement process, so that the time efficiency could be greatly improved. In order to verify the feasibility and effectiveness of the model, we conducted experiments and set up different methods as control experiments. The experimental result shows that the proposed H-LSTM method can not only make accurate predictions of future full-ocean deep SSP, but also accurately capture the periodic changes of SSP over time.

\section*{Conflict of Interest Statement}
The authors declare that the research was conducted in the absence of any commercial or financial relationships that could be construed as a potential conflict of interest.
\bibliographystyle{cas-model2-names}

\bibliography{my.bib}



%\nolinenumbers
\end{document}